\documentclass{elsart}

\usepackage{amsmath}
\usepackage{graphicx}
\usepackage{axodraw4j}

\def\phamp{2} 
\def\vtxsz{1}    

\def\HyyGeneric{\raisebox{-19pt}{\begin{picture}(101,44)(-12,-22)
  \Line[dash](0,0)(20,0)\Photon(60,20)(80,20){\phamp}{3}\Photon(60,-20)(80,-20){-\phamp}{3}
  \Line(20,  0)(60, 20)\Line(24, -2)(60, 16)\Line(28, -4)(60, 12)\Line(32, -6)(60,  8)\Line(36, -8)(60,  4)\Line(40,-10)(60,  0)\Line(44,-12)(60, -4)\Line(48,-14)(60, -8)\Line(52,-16)(60,-12)\Line(56,-18)(60,-16)\Line(60, 20)(60,-20)
  \Line(60,-20)(20,0)
  \Line[arrow,arrowpos=1,arrowwidth=0.9](64, 16)(75, 16)\Text(70, 13)[t]{$q_1$}
  \Line[arrow,arrowpos=1,arrowwidth=0.9](64,-16)(75,-16)\Text(70,-13)[b]{$q_2$}
  \Text(-2,0)[r]{$H$}\Text(82,20)[l]{$\mu$}\Text(82,-20)[l]{$\nu$}
\end{picture}}}

\def\HyyLO{\raisebox{-19pt}{\begin{picture}(101,44)(-12,-22)
  \Line[dash](0,0)(20,0)\Photon(60,20)(80,20){\phamp}{3}\Photon(60,-20)(80,-20){-\phamp}{3}
  \Line[arrow](20,0)(60,20)\Line(60,20)(60,-20)\Line(60,-20)(20,0)
  \Text(-2,0)[r]{$H$}\Text(82,20)[l]{$\gamma$}\Text(82,-20)[l]{$\gamma$}
\end{picture}}}
\def\HyyNLO{\raisebox{-19pt}{\begin{picture}(101,44)(-12,-22)
  \Line[dash](0,0)(20,0)\Photon(60,20)(80,20){\phamp}{3}\Photon(60,-20)(80,-20){-\phamp}{3}
  \Line[arrow](20,0)(60,20)\Line(60,20)(60,-20)\Line(60,-20)(20,0)
  \Gluon(40,-10)(60,0){\phamp}{3}
  \Text(-2,0)[r]{$H$}\Text(82,20)[l]{$\gamma$}\Text(82,-20)[l]{$\gamma$}
\end{picture}}}
\def\HyySA{\raisebox{-17pt}{\begin{picture}(121,44)(-12,-2)
  \Line[dash](0,20)(20,20)\Gluon(60,0)(40,0){\phamp}{3}\Gluon(40,40)(60,40){\phamp}{3}\Photon(80,0)(100,0){-\phamp}{3}\Photon(80,40)(100,40){\phamp}{3}\Text(-2,20)[r]{$H$}\Text(102,0)[l]{$\gamma$}\Text(102,40)[l]{$\gamma$}
  \Line[arrow](20,20)(40,40)\Line(40,40)(40,0)\Line(40,0)(20,20)
  \Line(60,0)(80,0)\Line[arrow](60,40)(80,40)\Line(60,0)(60,40)\Line(80,0)(80,40)\Text(32,20)[]{$t$}\Text(70,20)[]{$q$}
\end{picture}}}
\def\HyySB{\raisebox{-17pt}{\begin{picture}(121,44)(-12,-2)
  \Line[dash](0,20)(20,20)\Gluon(60,0)(40,0){\phamp}{3}\Gluon(40,40)(60,40){\phamp}{3}\Photon(80,0)(100,0){-\phamp}{3}\Photon(80,40)(100,40){\phamp}{3}\Text(-2,20)[r]{$H$}\Text(102,0)[l]{$\gamma$}\Text(102,40)[l]{$\gamma$}
  \Line[arrow](20,20)(40,40)\Line(40,40)(40,0)\Line(40,0)(20,20)
  \Line(60,0)(80,0)\Line[arrow](60,40)(80,40)\Line(60,0)(80,40)\Line(80,0)(60,40)\Text(32,20)[]{$t$}\Text(70,33)[]{$q$}
\end{picture}}}
\def\HyyNA{\raisebox{-19pt}{\begin{picture}(101,44)(-12,-22)
  \Line[dash](0,0)(20,0)\Photon(60,20)(80,20){\phamp}{3}\Photon(60,-20)(80,-20){-\phamp}{3}
  \Line[arrow](20,0)(60,20)\Line(60,20)(60,-20)\Line(60,-20)(20,0)
  \Gluon(33.33,-6.67)(60,6.67){\phamp}{4}\Gluon(46.67,-13.33)(60,-6.67){\phamp}{2}
  \Text(-2,0)[r]{$H$}\Text(82,20)[l]{$\gamma$}\Text(82,-20)[l]{$\gamma$}
\end{picture}}}
\def\HyyNB{\raisebox{-19pt}{\begin{picture}(101,44)(-12,-22)
  \Line[dash](0,0)(20,0)\Photon(60,20)(80,20){\phamp}{3}\Photon(60,-20)(80,-20){-\phamp}{3}
  \Line[arrow,arrowpos=0.25](20,0)(60,20)\Line(60,20)(60,-20)\Line(60,-20)(20,0)
  \Gluon(40,-10)(47.5,0){\phamp}{2}\Gluon(40,10)(47.5,0){\phamp}{2}\Gluon(60,0)(47.5,0){\phamp}{2}
  \Text(-2,0)[r]{$H$}\Text(82,20)[l]{$\gamma$}\Text(82,-20)[l]{$\gamma$}
\end{picture}}}
\def\HyyNC{\raisebox{-19pt}{\begin{picture}(101,44)(-12,-22)
  \Line[dash](0,0)(20,0)\Photon(60,20)(80,20){\phamp}{3}\Photon(60,-20)(80,-20){-\phamp}{3}
  \Line[arrow,arrowpos=0.25](20,0)(60,20)\Line(60,20)(60,-20)\Line(60,-20)(20,0)
  \Gluon(47.5,-5)(47.5,-13.75){\phamp}{1}\Gluon(47.5,13.75)(47.5,5){\phamp}{1}
  \BCirc(47.5,0){5}
  \Text(-2,0)[r]{$H$}\Text(82,20)[l]{$\gamma$}\Text(82,-20)[l]{$\gamma$}
\end{picture}}}

\def\LeftBubble{\Arc(-25,8)(5,0,240)\Line(-20,0)(-20,8)\Line(-20,0)(-27.5,3.67)\Vertex(-20,0){\vtxsz}}
\def\LeftLowerBubble{\Arc(-25,-8)(5,-240,0)\Line(-20,0)(-20,-8)\Line(-20,0)(-27.5,-3.67)\Vertex(-20,0){\vtxsz}}
\def\Ja{\begin{picture}(60,40)(-30,-20)\Line[dash](-30,0)(-20,0)\Line[dash](20,0)(30,10)\Line[dash](20,0)(30,-10)\LeftBubble\LeftLowerBubble
  \Arc[dash](0,0)(20,0,360)\Vertex(-20,0){\vtxsz}\Vertex(20,0){\vtxsz}
\end{picture}}
\def\Jdb{\begin{picture}(60,40)(-30,-20)\Line[dash](-30,0)(-20,0)\Line[dash](20,0)(30,10)\Line[dash](20,0)(30,-10)\LeftBubble
  \Arc[dash](-10,0)(10,0,360)\Arc[dash](10,0)(10,0,360)\Vertex(-20,0){\vtxsz}\Vertex(0,0){\vtxsz}\Vertex(20,0){\vtxsz}
\end{picture}}
\def\Jc{\begin{picture}(60,40)(-30,-20)\Line[dash](-30,0)(-20,0)\Line[dash](20,0)(30,10)\Line[dash](20,0)(30,-10)\LeftBubble
  \Line[dash](-20,0)(20,0)\Arc[dash](0,0)(20,0,360)\Vertex(-20,0){\vtxsz}\Vertex(20,0){\vtxsz}
\end{picture}}
\def\Jda{\begin{picture}(60,40)(-30,-20)\Line[dash](-30,0)(-20,0)\Line[dash](20,20)(30,20)\Line[dash](20,-20)(30,-20)
  \Line[dash](-20,0)(20,20)\Line[dash](-20,0)(20,-20)\Arc[dash](-14.64,0)(40,-30,30)\Arc[dash](54.64,0)(40,150,210)
  \Vertex(-20,0){\vtxsz}\Vertex(20,20){\vtxsz}\Vertex(20,-20){\vtxsz}\LeftBubble
\end{picture}}
\def\Jf{\begin{picture}(60,40)(-30,-20)\Line[dash](-30,0)(-20,0)\Line[dash](20,20)(30,20)\Line[dash](20,-20)(30,-20)
  \Line[dash](-20,0)(0,20)\Line[dash](-20,0)(0,-20)\Line[dash](20,20)(0,20)\Line[dash](20,-20)(0,-20)\Line[dash](20,-20)(0,20)\Line[dash](20,20)(0,-20)
  \Vertex(-20,0){\vtxsz}\Vertex(20,20){\vtxsz}\Vertex(20,-20){\vtxsz}\Vertex(0,20){\vtxsz}\Vertex(0,-20){\vtxsz}\LeftBubble
\end{picture}}
\def\Bca{\begin{picture}(60,40)(-10,0)
  \Arc(3.83,6)(6,101.25,318.75)\Arc(36.17,6)(6,-138.75,78.75)\Arc(20,34)(6,-18.75,198.75)\Line(20,15.33)(14.32,32.07)\Line(20,15.33)(25.68,32.07)\Line(20,15.33)(8.35,2.04)\Line(20,15.33)(2.66,11.88)\Line(20,15.33)(37.34,11.88)\Line(20,15.33)(31.65,2.04)\Vertex(20,15.33){\vtxsz}
\end{picture}}
\def\Bcia{\begin{picture}(60,40)(-30,-20)
  \Arc(0,0)(20,0,360)\Arc[dash](26.67,0)(33.33,143.13,216.87)\Arc[dash](-26.67,0)(33.33,-36.87,36.87)\Vertex(0,-20){\vtxsz}\Vertex(0,20){\vtxsz}
\end{picture}}
\def\Bcib{\begin{picture}(60,40)(-30,-20)
  \Arc(0,0)(20,0,360)\Arc(26.67,0)(33.33,143.13,216.87)\Arc(-26.67,0)(33.33,-36.87,36.87)\Vertex(0,-20){\vtxsz}\Vertex(0,20){\vtxsz}
\end{picture}}

\def\toposep{~~}

\begin{document}
\newcommand{\GeV}{\ensuremath{\mathrm{\,GeV}}}

\begin{frontmatter}
\begin{flushleft}
TTP12-046\\
SFB/CPP-12-94\\
ZU-TH 25/12\\
LPN12-130
\end{flushleft}
\title{Complete three-loop QCD corrections\\to the decay $H\to \gamma\gamma$}
\author[Zurich]{P.~Maierh\"ofer},
\author[Karlsruhe]{P.~Marquard}
\address[Zurich]{Institut f\"ur Theoretische Physik, Universit\"at Z\"urich, 8057 Z\"urich, Switzerland}
\address[Karlsruhe]{Institut f\"ur Theoretische Teilchenphysik,
 Karlsruhe~Institute~of~Technology~(KIT), 76128 Karlsruhe, Germany}
\begin{abstract}
  We present the result for the three-loop singlet QCD corrections to
  the decay of a Higgs boson into two photons and improve the calculation
  for the non-singlet case. With the new result
  presented, the decay width $\Gamma(H \to \gamma\gamma )$ is completely
  known at ${\cal O}(G_F \alpha^2 \alpha_s^2, G_F \alpha^3)$.
\end{abstract}
\end{frontmatter}

\section{Introduction}
\label{sec:introduction}

After the discovery of a new boson at about 126 GeV at the LHC
\cite{ATLAS:2012ae,Chatrchyan:2012tx} it is imperative to study all
decay modes of the Standard Model Higgs boson as precisely as
possible. In this paper we are concerned with the decay of the Higgs
boson into two photons which is the decay channel with the largest
significance for the discovery of the Higgs boson at the LHC.

The decay of a Higgs boson into two photons is 
mediated through either charged gauge bosons or top quarks at
one-loop order.  The decay rate can be cast into the form
\begin{equation}
  \label{eq:1}
  \Gamma(H \to \gamma\gamma) = \frac{M_H^3}{64 \pi}|A_W + A_t |^2\,,
\end{equation}
with the leading order values
\begin{equation}
\begin{split}
  \label{eq:2}
  &A_W^{(0)} = -\frac{\alpha \sqrt{\sqrt{2} G_F}}{2 \pi} \left( 2 + \frac{3}{\tau_W} + \frac{3}{\tau_W} \left (2 - \frac{1}{\tau_W} \right )\arcsin^2 \sqrt{\tau_W}\right ), \\
  &A_t^{(0)} = {\hat A}_t \frac{3}{2 \tau}\left( 1 + \left (1 - \frac{1}{\tau} \right)\arcsin^2 \sqrt{\tau}\right ),\\
  &{\hat A}_t = N_c\frac{ 2 \alpha \sqrt{\sqrt{2} G_F}}{3 \pi} Q_t^2,
\end{split}
\end{equation}
where $\tau_W = M_H^2 / (4 M_W^2)$ and $\tau = M_H^2 / (4 m_t^2)$.  The
former contribution is larger than the latter (by a factor 4.5) and
opposite in sign. Both contributions have been investigated in great
detail in the literature. The two-loop QCD corrections to the decay have
first been evaluated in the heavy-top limit in
Refs.~\cite{Zheng:1990qa,Djouadi:1990aj,Dawson:1992cy} and later, keeping
the full top-mass dependence, in
Refs.~\cite{Fleischer:2004vb,Harlander:2005rq,Aglietti:2006tp}.  The
two-loop electroweak corrections have been investigated
in~\cite{Actis:2008ts,Passarino:2007fp,Degrassi:2005mc,Fugel:2004ug}. Combining
the two-loop QCD and electroweak corrections one observes a nearly
complete cancellation between these two contributions for $M_H = 126
\GeV$ as discussed below.

\begin{figure}[t]
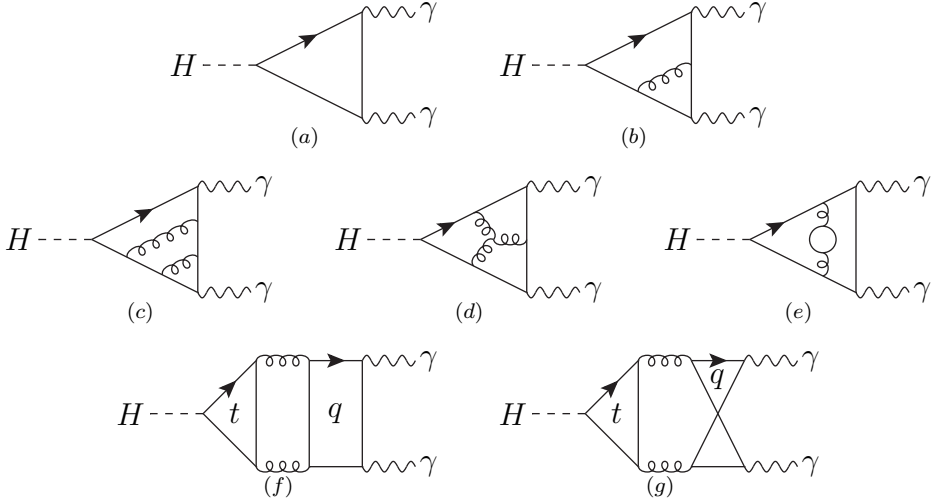

  \begin{center}
    $\underset{(a)}{\HyyLO}\qquad\underset{(b)}{\HyyNLO}$\\[2ex]
    $\underset{(c)}{\HyyNA}\qquad\underset{(d)}{\HyyNB}\qquad\underset{(e)}{\HyyNC}$\\[2ex]
    $\underset{(f)}{\HyySA}\qquad\underset{(g)}{\HyySB}$\\
  \end{center}
  \caption{Sample diagrams of the 1-loop (a), 2-loop (b), 3-loop non-singlet (c)-(e), and 3-loop singlet (f)-(g) top quark induced contribution to $H\to\gamma\gamma$.}
  \label{SampleHyyDiagrams}
\end{figure}

At next-to-next-to-leading order (NNLO) the non-singlet QCD
contributions (cf Fig.~\ref{SampleHyyDiagrams} (c)-(e)) have been
calculated in the heavy top limit, including additional terms in an
expansion in $\tau = M_H^2/(4 m_t^2)$, in
Ref.~\cite{Steinhauser:1996wy}. At this order a new class of diagrams,
so-called singlet diagrams (cf Fig.~\ref{SampleHyyDiagrams} (f)-(g)),
contributes for the first time. They can be characterized by the
property, that the external lines are coupled to different fermion
loops. The contribution from this kind of diagrams has not been taken
into account up to now and there is no formal argument, that it should
be suppressed compared to the non-singlet contribution, when
considering the same order in $\alpha_s$. In this letter we present
the calculation of this last missing piece to obtain a complete NNLO
QCD prediction for the decay of a Higgs boson into two photons. In
addition, to improve the existing prediction and to check our setup, we
recalculated the non-singlet contribution and added more terms in the
expansion in $\tau$.

This letter is organized as follows: In Section~\ref{sec:calculation}
we describe the necessary steps of the calculation and in Section
\ref{sec:results} we present the full three-loop QCD corrections to
the decay width. In Section \ref{sec:conclusion} we give the numerical
result and our conclusions.

\section{Calculation}
\label{sec:calculation}

We start from the decay amplitude of a Higgs boson into two photons
\begin{align}
  \mathcal{A}^{\mu\nu} = \HyyGeneric.
\end{align}
Decomposing the Lorentz structure in the most general way the amplitude $\mathcal{A}^{\mu\nu}$ can be written as
\begin{align}
  \mathcal{A}^{\mu\nu} = q_1^\nu q_2^\mu A + g^{\mu\nu} B + q_1^\mu
  q_2^\nu C + q_1^\mu q_1^\nu D + q_2^\mu q_2^\nu E\,,
\end{align}
with scalar form factors $A$, $B$, $C$, $D$, and $E$, which can be calculated by applying the projectors
\begin{align}
  P_A^{\mu\nu} &= \frac{(d-1)q_1^\mu q_2^\nu + q_1^\nu q_2^\mu - q_1q_2 g^{\mu\nu}}{(d-2)(q_1q_2)^2}, \nonumber\\
  P_B^{\mu\nu} &= \frac{q_1q_2 g^{\mu\nu} - q_1^\mu q_2^\nu - q_1^\nu q_2^\mu}{(d-2)q_1q_2}, \nonumber\\
  P_C^{\mu\nu} &= \frac{(d-1) q_1^\nu q_2^\mu + q_1^\mu q_2^\nu - q_1q_2 g^{\mu\nu}}{(d-2)(q_1q_2)^2},\\
  P_D^{\mu\nu} &= \frac{q_2^\mu q_2^\nu}{(q_1q_2)^2}, \nonumber\\
  P_E^{\mu\nu} &= \frac{q_1^\mu q_1^\nu}{(q_1q_2)^2} \nonumber
\end{align}
on the amplitude (e.\,g. $A=P_A^{\mu\nu}\mathcal{A}_{\mu\nu}$). Due to
gauge invariance the contractions $q_{1\mu}\mathcal{A}^{\mu\nu}$ and
$q_{2\nu}\mathcal{A}^{\mu\nu}$ of the amplitude with the momenta of the
external photons vanish. This imposes the relations $B=-q_1q_2A$ and
$D=E=0$ on the form factors. In order to verify these relations
explicitly we calculate $A$, \dots, $E$ separately. The coefficient $C$,
although non zero, does not contribute to the decay rate of the Higgs boson because of the transversality of the photon wave functions. Therefore the Lorentz structure of the physical amplitude simplifies to
\begin{align}
  \mathcal{A}^{\mu\nu} = (q_1^\nu q_2^\mu - q_1q_2\,g^{\mu\nu}) A.
\end{align}
To obtain the decay width, the combination $A=A_W+A_t$ is inserted into
eq.~(\ref{eq:1}) for the $W$ and the top quark loop induced
contributions.

Since the calculation of the decay amplitude including the full
dependence on Higgs-boson and top-quark mass is not possible at the
moment, only an expansion in $\tau = M_H^2 / (4 m_t^2)$ can be
obtained. The expansion of the amplitude in $\tau$ can be performed
following two conceptually different strategies. The first option, which
is only applicable for the calculation of a few terms in the expansion,
relies on the direct expansion of each Feynman diagram in $\tau$ and the
subsequent reduction of the appearing integrals to master
integrals. Since this method becomes very time-consuming when
calculating higher terms in the expansion, a second approach has been
devised. In this alternative approach, the full mass dependence is kept
as long as possible, meaning that the reduction to master integrals is
performed keeping the full mass dependence and only the master integrals
 are calculated in an expansion in $\tau$. In the following we
sketch the steps necessary to perform the calculation and explain how
the results for the needed master integrals in an expansion in $\tau$
can be obtained.

\begin{figure}
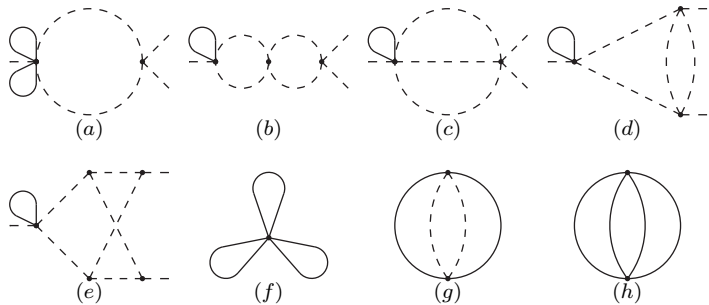

  \begin{center}
    $\underset{(a)}{\Ja}$\toposep$\underset{(b)}{\Jdb}$\toposep$\underset{(c)}{\Jc}$\toposep$\underset{(d)}{\Jda}$\\[2ex]
    $\underset{(e)}{\Jf}$\toposep$\underset{(f)}{\Bca}$\toposep$\underset{(g)}{\Bcia}$\toposep$\underset{(h)}{\Bcib}$
  \end{center}
  \caption{Integrals which serve as boundary conditions for the
    solutions of the differential equations for the vertex master
    integrals. Dashed internal lines are massless, solid lines carry the
    mass $m_t$. External lines on the right side of the diagrams (a)-(e)
    are massless on-shell, the external line on the left side is
    on-shell with mass $M_H$. These diagrams only contribute to the
    singlet case.}
  \label{boundaries}
\end{figure}

The Feynman diagrams are generated with
\texttt{qgraf}~\cite{Nogueira:1991ex} and processed by a
\texttt{Mathematica} program to map the integrals on $4$ singlet and
$15$ non-singlet topologies. A sample of the contributing diagrams is
depicted in Fig.~\ref{SampleHyyDiagrams}. All appearing integrals are
reduced to master integrals using \texttt{Crusher}~\cite{crusher}. To
expand the master integrals $M_i(\tau)$ in $\tau=M_H^2/(4m_t^2)$ a
series ansatz
\begin{align}
  M_i(\tau) = \sum_k \big(M_{i,0}^{(k)} + M_{i,1}^{(k)} (-\tau)^{-\epsilon} + M_{i,2}^{(k)} (-\tau)^{-2\epsilon}\big) \tau^k
\end{align}
is inserted into the differential equation \cite{Remiddi:1997ny}
\begin{align}
  \Big(M_H^2\frac{\partial}{\partial M_H^2} + m_t^2\frac{\partial}{\partial
    m_t^2} - \frac{1}{2}\hat{D}\Big) M_i(\tau) = 0,
\end{align}
where $\hat{D}$ applied to $M_i(\tau)$ returns the mass dimension of
$M_i$. The differential equation follows from the fact that $M_i(\tau)$
is a homogeneous function in its dimensionful parameters $M_H^2$ and
$m_t^2$. Quarks except for the top quark are treated as massless. Using
the series ansatz results in a system of algebraic equations which are
systematically solved for the coefficients $M_{i,j}^{(k)}$ by a
\texttt{Mathematica} program which uses \texttt{Fermat} \cite{fermat} to simplify
rational functions. By this procedure all coefficients $M_{i,j}^{(k)}$
are expressed as linear combinations of the integrals which are depicted
in Fig.~\ref{boundaries} and serve as boundary conditions. Coefficients
$M_{i,1}^{(k)}$ and $M_{i,2}^{(k)}$ of non-integer powers of $\tau$ arise
only in the singlet contribution due to the massless cuts of the singlet
diagrams~\cite{Maier:2007yn}.

\section{Results}
\label{sec:results}

The top quark loop induced amplitude $A_t$ is expressed as a perturbative series in the strong coupling constant $\alpha_s$ with the one-, two-, and three-loop contributions $A_t^{(0)}$, $A_t^{(1)}$, and $A_t^{(2)}$,
\begin{align}
  A_t = \hat{A}_t~\Big[~A_t^{(0)} + \frac{\alpha_s}{\pi} A_t^{(1)} + \Big(\frac{\alpha_s}{\pi}\Big)^2 A_t^{(2)} + \dots \Big].
\end{align}
The three-loop contribution is, furthermore, split into the non-singlet part $A_{t,0}^{(2)}$, the singlet part with two top-quark loops $A_{t,t}^{(2)}$, and the singlet part with a top-quark and light-quark loop $A_{t,q}^{(2)}$
\begin{align}
  A_t^{(2)} = A_{t,0}^{(2)} + A_{t,t}^{(2)} + (Q_t^{-2}\sum_{q\neq t} Q_q^2) A_{t,q}^{(2)}
\end{align}
$Q_t$ and $Q_q$ are the electromagnetic charges of the top quark and the light quarks ($q\in \{u,d,s,c,b\}$).

The full results with the expansion in $\tau$ up to $\tau^{20}$
including generic $SU(N)$ colour factors and renormalisation scale
dependence are available
online.\footnote{\texttt{http://www-ttp.particle.uni-karlsruhe.de/Progdata/ttp12/ttp12-046/}} For
the presentation of the results in this section we truncate the
expansion after $\tau^{5}$
and insert the $SU(3)$ colour factors explicitly. The top-quark mass
is renormalized in the $\overline{\mathrm{MS}}$ scheme. We use the
abbreviations $L_\mu = \log{\mu^2 / m_t^2}$ in the
non-singlet and $L_\tau = \log (- 4 \tau )$ in the singlet
contribution.  For completeness also the $\alpha_s^0$ and $\alpha_s^1$
contributions are given as an expansion in $\tau$. 

\begin{figure}
  \includegraphics[width=0.9\linewidth]{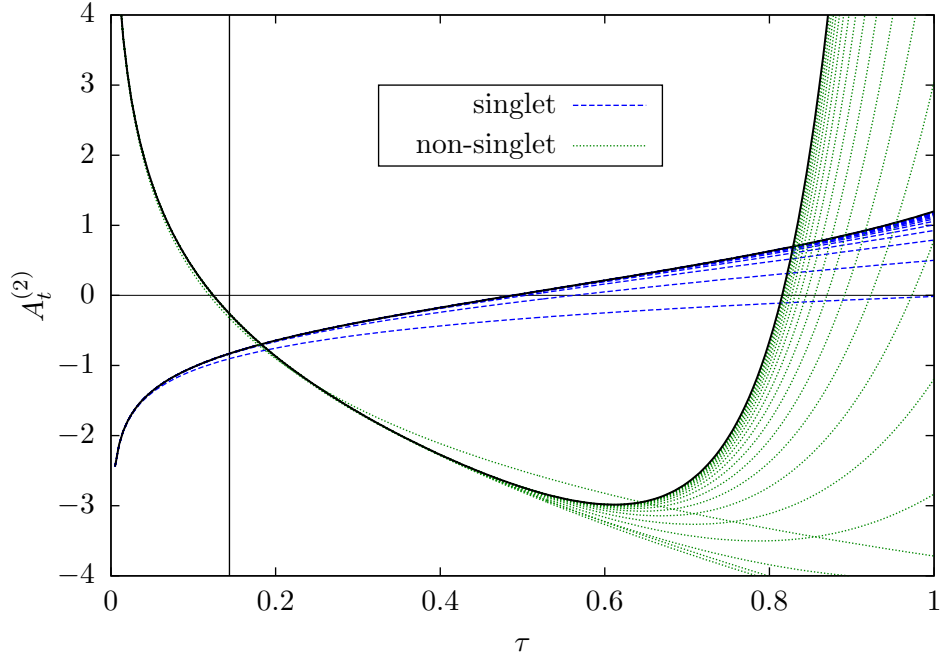}
  \caption{Numerical results for the non-singlet and singlet
    contribution at three-loop order. In the plot the first 20 terms in
    the expansion in $\tau$ are shown. The singlet plot shows the sum
    $A^{(2)}_{t,t}+\sum_{q\neq t}Q_q^2/Q_t^2A^{(2)}_{t,q}$. For the plot
    of the non-singlet contribution $A^{(2)}_{t,0}$ we set the
    renormalization scale $\mu = M_H$. The vertical line at $\tau=0.14$
    marks the value of $\tau$ for $M_H=126\GeV$. The convergence of the
    expansion for $\tau \to 1 $ can be improved by using the on-shell
    scheme for the top-quark mass.}
  \label{fig:3}
\end{figure}

{\allowdisplaybreaks

\begin{align}
  A_t^{(0)} =
   &+1
    +\tau\Big(\frac{7}{30}\Big)
    +\tau^2\Big(\frac{2}{21}\Big)
    +\tau^3\Big(\frac{26}{525}\Big)
    +\tau^4\Big(\frac{512}{17325}\Big)
    +\tau^5\Big(\frac{1216}{63063}\Big)\\[2ex]
  A_t^{(1)} =
   &-1
    +\tau\Big(\frac{38}{135}-\frac{7}{15}L_{\mu}\Big)
    +\tau^2\Big(\frac{1664}{14175}-\frac{8}{21}L_{\mu}\Big)\nonumber\\
   &+\tau^3\Big(\frac{12626}{496125}-\frac{52}{175}L_{\mu}\Big)
    +\tau^4\Big(-\frac{40664}{2338875}-\frac{4096}{17325}L_{\mu}\Big)\nonumber\\
   &+\tau^5\Big(-\frac{9128671204}{245827456875}-\frac{12160}{63063}L_{\mu}\Big)\\[2ex]
  A_{t,0}^{(2)} =
   &-\frac{31}{24}
    -\frac{7}{4}L_{\mu}\nonumber\\
   &+\tau\Big(
      -\frac{22326329}{622080}
      +\frac{4116067}{138240}\zeta_3
      -\frac{769}{1080}L_{\mu}
      +\frac{7}{120}L_{\mu}^2\Big)\nonumber\\
   &+\tau^2\Big(
      -\frac{68094821183}{2612736000}
      +\frac{508541309}{23224320}\zeta_3
      -\frac{1241}{1575}L_{\mu}
      +\frac{3}{7}L_{\mu}^2\Big)\nonumber\\
   &+\tau^3\Big(
      -\frac{102458003430188113}{122903101440000}
      +\frac{190929277363}{275251200}\zeta_3\nonumber\\
     &\hspace*{6ex}-\frac{256363}{496125}L_{\mu}
      +\frac{221}{350}L_{\mu}^2\Big)\nonumber\\
   &+\tau^4\Big(
      -\frac{489470768471800920451}{202790117376000000}
      +\frac{547186023461087}{272498688000}\zeta_3\nonumber\\
     &\hspace*{6ex}-\frac{14462}{66825}L_{\mu}
      +\frac{512}{693}L_{\mu}^2\Big)\nonumber\\
   &+\tau^5\Big(
      -\frac{114742890543235030923359893}{6430815309390151680000}
      +\frac{1766782778485181879}{119027426918400}\zeta_3\nonumber\\
     &\hspace*{6ex}+\frac{307117801}{7449316875}L_{\mu}
      +\frac{1520}{1911}L_{\mu}^2\Big)\\[2ex]
  A_{t,t}^{(2)} =
   &+\frac{1}{8}\nonumber\\
   &+\tau\Big(
      -\frac{28777}{207360}
      +\frac{749}{3072}\zeta_3\Big)\nonumber\\
   &+\tau^2\Big(
      +\frac{34183679}{522547200}
      +\frac{18935}{221184}\zeta_3
      -\frac{2}{45}L_\tau\Big)\nonumber\\
   &+\tau^3\Big(
      -\frac{2665898377}{390168576000}
      +\frac{1419929}{23592960}\zeta_3
      -\frac{41}{2025}L_\tau\Big)\nonumber\\
   &+\tau^4\Big(
      -\frac{37746440955049}{1931334451200000}
      +\frac{11964631}{235929600}\zeta_3
      -\frac{598}{42525}L_\tau\Big)\nonumber\\
   &+\tau^5\Big(
      -\frac{2381500300071333647}{231991894278144000000}
      +\frac{862749257}{28311552000}\zeta_3
      -\frac{74924}{7016625}L_\tau\Big)\\[2ex]
  A_{t,q}^{(2)} =
   &-\frac{13}{12}
      +\frac{2}{3}\zeta_3
      +\frac{1}{6}L_\tau\nonumber\\
   &+\tau\Big(
      -\frac{3493}{48600}
      +\frac{7}{45}\zeta_3
      +\frac{19}{1620}L_\tau\Big)\nonumber\\
   &+\tau^2\Big(
      -\frac{3953}{396900}
      +\frac{4}{63}\zeta_3
      -\frac{1}{3780}L_\tau\Big)\nonumber\\
   &+\tau^3\Big(
      -\frac{3668899}{2679075000}
      +\frac{52}{1575}\zeta_3
      -\frac{1696}{1063125}L_\tau\Big)\nonumber\\
   &+\tau^4\Big(
      +\frac{414962}{2210236875}
      +\frac{1024}{51975}\zeta_3
      -\frac{136}{91125}L_\tau\Big)\nonumber\\
   &+\tau^5\Big(
      +\frac{611578464557}{1409328810264375}
      +\frac{2432}{189189}\zeta_3
      -\frac{7571576}{6257426175}L_\tau\Big)
\end{align}

} 
The results for $A_{t,0}^{(2)}$ agree with the results presented in
Ref.~\cite{Steinhauser:1996wy}.  In Tab.~\ref{tab:1} we give the results
for the first 20 coefficients of the series in $\tau$ in numerical
form. For the singlet contribution we show the results for the constant
and the logarithmic part proportional to $L_\tau$ separately.  To
illustrate the convergence of the series we show all 20 terms of the
expansion of the three-loop contribution in graphical form in
Fig.~\ref{fig:3}. Around $\tau = 0.14$, corresponding to the Higgs-boson
mass favoured by the LHC experiments, the first four (five) terms in the
expansion are sufficient to obtain an accurate result with a relative
error of $10^{-5}$ wrt.\ the total 3-loop amplitude for the singlet (non-singlet)
contribution.  At larger values of $\tau$  more terms of the
expansion have to be taken into account, e.g. at  $\tau = 0.5$ nine (sixteen)
terms are needed to obtain similar accuracy.

Collecting all available information the amplitude can be cast in the form
\begin{equation}
A_{H \to \gamma\gamma } = \underbrace{A_{W}^{(0)} +
  \hat A_t A_t^{(0)}}_{A_{\mathrm{LO}}}
+ \frac{\alpha}{\pi} \underbrace{ 
    A_{\mathrm{EW}}^{(1)} }_{A_{\mathrm{NLO-EW}}}
+ \frac{\alpha_s}{\pi} \underbrace{  \hat A_t A_{t}^{(1)} }_{A_{\mathrm{NLO-QCD}}}
+ \left(
  \frac{\alpha_s}{\pi}\right)^2 \underbrace{ \hat A_t A_{t}^{(2)}}_{A_{\mathrm{NNLO}}} \,,
\end{equation}
where $A_{\mathrm{EW}}^{(1)}$ denotes the electroweak corrections to the
$W$ and top-quark induced processes combined.
The partial decay width is then given by
\begin{align}
\Gamma_{H \to \gamma\gamma } = \frac{M_H^3}{64 \pi} \big ( A^2_{\mathrm{LO}} +
\frac{\alpha}{\pi} ( 2 A_{\mathrm{LO}} A_{\mathrm{NLO-EW}} ) +
\frac{\alpha_s}{\pi} ( 2 A_{\mathrm{LO}} A_{\mathrm{NLO-QCD}}) \nonumber \\ +
\left ( \frac{\alpha_s}{\pi} \right )^2 (2 A_{\mathrm{LO}}
\mathrm{Re} ( A_{\mathrm{NNLO}} ) + A^2_{\mathrm{NLO}}) \big ) \label{eq:3}
\,.
\end{align}
For a Higgs boson with a mass of $M_H = 126\GeV$ ($\tau = 0.14$) this
evaluates numerically to
\begin{align}
  \Gamma_{H\to\gamma\gamma} &= (
    9.398\cdot 10^{-6}
   -1.48\cdot 10^{-7}
   +1.68\cdot 10^{-7}
   +7.93\cdot 10^{-9}
  ) \GeV \\ \nonumber
  &= 9.425\cdot 10^{-6}\GeV,
\end{align}
where we used $m_t(M_H) = 166 \GeV$\footnote{The
  $\overline{\mathrm{MS}}$ mass $m_t(M_H)$ was obtained from the
  on-shell mass $M_t=172.64\GeV$ using the
  $\overline{\mathrm{MS}}$-on-shell relation at 3-loop accuracy \cite{Melnikov:2000qh,Marquard:2007uj,Melnikov:2000zc}.},
$\alpha_s(M_H) / \pi = 0.0358 $, $G_F= 1.16637\cdot 10^{-5} \GeV^{-2}$, and $\alpha =
\alpha(0) = 1/137$ as input parameters. The value for the two-loop
electroweak correction was taken from
Ref.~\cite{Passarino:2007fp}.  It has to be noted, that there is a
partial cancellation between the two-loop QCD and electroweak
corrections. To assess the influence of the singlet diagrams the
next-to-next-to-leading order term can be further decompsed as
\[
\Gamma_{H \to \gamma\gamma}|_{\mathrm{NNLO}} = ( 7.5 \cdot 10^{-10}
|_{\mathrm{NLO}^2} + 1.73 \cdot 10^{-9} |_{\mathrm{non-singlett}} + 5.45
\cdot 10^{-9} | _{\mathrm{singlet}} ) \GeV \,,
\]
which shows, that the singlet diagrams are a factor of three larger than
the non-singlet ones and thus the most important  three-loop
contribution. 

\begin{table}
  \begin{center}
    \begin{tabular}{|r|r|r|r|r|r|r|r|}\hline
      \multicolumn{1}{|c}{$n$} & \multicolumn{1}{|c}{$A_{t,0}^{(2)}$} & \multicolumn{1}{|c}{$A_{t,q}^{(2)}[0]$} & \multicolumn{1}{|c}{$A_{t,q}^{(2)}[L_\tau ]$} & \multicolumn{1}{|c}{$A_{t,t}^{(2)}[0]$} & \multicolumn{1}{|c|}{$A_{t,t}^{(2)}[L_\tau ]$} \\ \hline
       0 & -1.29166667 & -0.28196206 &  0.16666667 & 0.12500000 &  0.00000000 \\
       1 & -0.09881144 &  0.11511420 &  0.01172840 & 0.15430166 &  0.00000000 \\
       2 &  0.25870689 &  0.06636139 & -0.00026455 & 0.16832244 & -0.04444444 \\
       3 &  0.16373113 &  0.03831749 & -0.00159530 & 0.06551243 & -0.02024691 \\
       4 &  0.08688592 &  0.02387041 & -0.00149246 & 0.04141534 & -0.01406232 \\
       5 &  0.05709451 &  0.01588624 & -0.00121001 & 0.02636532 & -0.01067807 \\
       6 &  0.05798951 &  0.01114079 & -0.00095670 & 0.01555892 & -0.00622762 \\
       7 &  0.07545680 &  0.00814321 & -0.00075859 & 0.01272091 & -0.00587319 \\
       8 &  0.10086885 &  0.00615332 & -0.00060811 & 0.00751470 & -0.00333698 \\
       9 &  0.12935841 &  0.00477756 & -0.00049383 & 0.00708618 & -0.00353585 \\
      10 &  0.15827468 &  0.00379378 & -0.00040622 & 0.00423275 & -0.00200757 \\
      11 &  0.18622498 &  0.00307013 & -0.00033822 & 0.00437893 & -0.00229504 \\
      12 &  0.21252358 &  0.00252489 & -0.00028474 & 0.00264217 & -0.00130721 \\
      13 &  0.23687981 &  0.00210551 & -0.00024213 & 0.00291761 & -0.00157907 \\
      14 &  0.25922151 &  0.00177712 & -0.00020779 & 0.00177454 & -0.00090205 \\
      15 &  0.27959438 &  0.00151591 & -0.00017979 & 0.00205621 & -0.00113694 \\
      16 &  0.29810454 &  0.00130527 & -0.00015673 & 0.00125855 & -0.00065080 \\
      17 &  0.31488570 &  0.00113331 & -0.00013755 & 0.00151292 & -0.00084866 \\
      18 &  0.33008045 &  0.00099137 & -0.00012148 & 0.00093081 & -0.00048637 \\
      19 &  0.34383005 &  0.00087306 & -0.00010789 & 0.00115159 & -0.00065225 \\
      20 &  0.35626884 &  0.00077356 & -0.00009631 & 0.00071159 & -0.00037400 \\ \hline
    \end{tabular}
  \end{center}
  \caption{Numerical values of the first 20 terms in the expansion in
    $\tau$ for the three-loop non-singlet $A_{t,0}^{(2)}$ and singlet
    contribution $A_{t,t}^{(2)}$ and $A_{t,q}^{(2)}$. For the
    non-singlet part we set $\mu = m_t$. In case of the
    singlet contribution the results for the constant part ($A_{t}^{(2)}[0]$)
    and the logarithmic part ($A_{t}^{(2)}[L_\tau ]$) are shown
    separately.}
\label{tab:1}
\end{table}


\section{Conclusion}
\label{sec:conclusion}

We presented new results for the singlet diagrams which contribute to
the decay of a Higgs boson into two photons at next-to-next-to-leading
order, which up to now have not been considered. The corrections to the
decay rate due to singlet diagrams are about a factor of three larger
than the non-singlet ones.  An improved prediction for the non-singlet
contribution reduces the error of the three-loop contribution, which can
be neglected for a Higgs-boson mass of $126 \GeV$. The total partial
decay width is $\Gamma_{H\to\gamma\gamma} =
\Gamma_{\mathrm{LO}}+\Gamma_{\mathrm{NLO-EW}}+\Gamma_{\mathrm{NLO-QCD}}+\Gamma_{\mathrm{NNLO}}
= ( 9.398\cdot 10^{-6} -1.48\cdot 10^{-7} +1.68\cdot 10^{-7} +7.93\cdot
10^{-9} ) \GeV = 9.425\cdot 10^{-6}\GeV$.

\section*{Acknowledgements}

We are grateful to S.~Uccirati and C.~Sturm for providing the exact
numerical value for the two-loop electroweak contribution.  We like to
thank N.~Zerf for crosschecks on the non-singlet part and J.H.~K\"uhn,
M.~Steinhauser and T.~Kasprzik for carefully reading the
manuscript. This work was supported by the DFG through the SFB/TR 9
``Computational Particle Physics". The work of Ph.M. was supported by the
Swiss National Science Foundation.

\end{document}